\documentclass[floatfix,aps,amsmath,nofootinbib,onecolumn,10pt]{revtex4}
\pdfoutput=1
\usepackage{listings}
\usepackage{graphicx}
\usepackage{bm}
\usepackage{rotating}
\usepackage{array}
\usepackage{amsmath}
\usepackage{amssymb} 
\usepackage{mathrsfs} 
\usepackage{cancel}

\def\({\left(}
\def\){\right)}
\def\[{\left[}
\def\]{\right]}

\def\e{\begin{equation}}
\def\q{\end{equation}}
\def\m{\begin{eqnarray}}
\def\n{\end{eqnarray}}

\begin{document}

\title{Signatures of primordial gravitational waves in matter power spectrum }

\author{  Ke Wang$^1$ \footnote{wangke@itp.ac.cn}}
\affiliation{
$^1$ National Astronomical Observatories, \\Chinese Academy of Sciences,  20A Datun Road, Beijing 100012, China\\}

\date{\today}

\begin{abstract}

We simulate the evolution of a dust universe from $z=1089$ to $z=0$ by numerically integrating the Einstein's equation for a spatially flat Friedmann-Lemaire-Robertson-Walker (FLRW) background spacetime with scalar perturbations which are derived from the matter power spectrum produced with the Code for Anisotropies in the Microwave Background (CAMB). To investigate the effects of primordial gravitational waves (GWs) on the inhomogeneity of the universe, we add an additional decaying, divergenceless and traceless primordial tensor perturbation with its initial amplitude being $3\times 10^{-4}$ to the above metric.
We find that this primordial tensor perturbation suppresses the matter power spectrum by about $0.01\%$ at $z=0$ for modes with wave number similar to its.
This suppression may be a possible probe of a GWs background in the future.

\end{abstract}

\pacs{???}

\maketitle


\section{Introduction}
\label{introduction}

One of the most important predictions by inflation \cite{Starobinsky:1980te,Guth:1980zm} is that there is a stochastic gravitational waves (GWs) background. So far, people have made every endeavor to detect such a GWs background and test inflation scenario experimentally: the most promising one is the B-mode polarization of the cosmic microwave background (CMB) \cite{Seljak:1996gy,Kamionkowski:1996zd,Kamionkowski:2015yta}; the complementary and even more sensitive one is the 21cm HI emission from the dark ages \cite{Book:2011dz,Masui:2010cz}; some not very competitive ones including weak lensing shear \cite{Dodelson:2003bv,Dodelson:2010qu} and other large-scale structure observables \cite{Jeong:2012nu,Schmidt:2012nw}.
The goal of this paper is to investigate the signatures of primordial GWs in matter power spectrum with numerical relativity, thereby proposing a possible probe of a GWs background.

As we known, massive neutrinos will slow the gravitational collapse of halos  on scales smaller than their free-streaming length when they become non-relativistic, which will affect the way large-scale cosmological structures form and lead to a suppression in the galaxy power spectrum on small scales  observed today. Therefore, people can constrain the upper limit on the sum of neutrino masses from the power spectrum of galaxy surveys \cite{Hu:1997mj,Lesgourgues:2006nd,Riemer-Sorensen:2013jsa,Palanque-Delabrouille:2015pga,Cuesta:2015iho}. As for the matter power spectrum on large scales, it would not be modified significantly by radiation, neutrinos or baryons. So the matter power spectrum on large scales can serve as another handle on the primordial fluctuations and inflation.

So far, the power spectrum data from the Clustering of the Sloan Digital Sky Survey DR7 Luminous Red Galaxies ranges from $k=0.02 h \text{Mpc}^{-1}$ to $k=0.2 h \text{Mpc}^{-1}$ \cite{Reid:2009xm} and the power spectrum data from the WiggleZ Dark Energy Survey ranges from $k=0.01 h \text{Mpc}^{-1}$ to $k=0.5 h \text{Mpc}^{-1}$ \cite{Parkinson:2012vd}. Due to their small $k$ span, these data are not suitable to constrain the large-scale primordial fluctuations and inflation. However, the future high precision lensing and galaxy redshift surveys, such as the Large Synoptic Survey Telescope (LSST) \cite{lsst,Abell:2009aa}, will has a large enough $k$ span to confirm the turnover in the power spectrum and constrain the large-scale primordial fluctuations. So, in this paper, we will consider primordial tensor perturbations with comparable wave number to the scale of turnover.

Here, our work is based on the wide-used \textsf{Einstein Toolkit} \cite{Loffler:2011ay} to integrate Einstein's equation: the thorn \textsf{ML\_BSSN} \cite{Brown:2008sb,Reisswig:2010cd,code} was used to evolve spacetime using the Baumgarte-Shapiro-Shibata-Nakamura (BSSN) formalism \cite{Baumgarte:1998te,Shibata:1995we,Alcubierre:2000xu} and the thorn \textsf{GRHydro} was used to evolve the hydrodynamical system \cite{Moesta:2013dna,Baiotti:2004wn,Hawke:2005zw}. Moreover, we initialize an almost FLRW Universe with scalar and tensor perturbations as \cite{Macpherson:2016ict,Wang:2018qfr}, and especially turn to the matter power spectrum as \cite{Macpherson:2018btl}.

This paper is organized as follows. In Sec.~\ref{initial}, we give the initial conditions for background by rescaling the scale factor and perturbations by analyzing the matter power spectrum. In Sec.~\ref{results}, we show the results of simulations. At last, a brief summary and discussion are included in Sec.~\ref{summary}.
In this paper, we adopt the following conventions: Greek indices run in \{0, 1, 2, 3\}, Latin indices run in \{1, 2, 3\} and repeated indices implies summation.

\section{Initial conditions}
\label{initial}

\subsection{Initial conditions for background}
\label{background}

Since we will perform large-scale cosmological simulations instead of the simulations of black-hole-binary-like astrophysical system,
we modify the file \textsf{EOS\_Omni\_Module.F90} in \textsf{Einstein Toolkit} to replace the default unit system: $\text{M}_{\odot}=\text{G}=\text{c}=1$ with the new one: $1\text{Mpc}=\text{G}=\text{c}=1$ \cite{Macpherson:2018btl}. Under this new unit system and with the cosmological parameters consistent with Planck 2018 results \cite{Aghanim:2018eyx} as shown in Tab.~\ref{tab:parameter}, the matter density of our universe is $\bar{\rho}_m^P=6.0\times10^{-9}\times0.3166$ at $z=0$, hence $\bar{\rho}_m^P=6.0\times10^{-6}\times0.3166$ at $z=9$. Considering a fiducial universe whose matter density $\bar{\rho}_m^F(z)$ is equal to $\bar{\rho}_m^P(z)$, the scale factor of this fiducial one $a^F=10a^P$ as shown in Tab.~\ref{tab:a} means that the comoving matter density of it is $\bar{\rho}_m^F*=6.0\times10^{-6}\times0.3166$ as shown in Tab.~\ref{tab:comoving}. Here we will turn to a blown-up fiducial universe by 109 times to mimic our universe in simulations: setting the scale factor used during simulations as $a^S=109a^F$ as shown in Tab.~\ref{tab:a} and keeping the comoving matter density being $\bar{\rho}_m^S*=6.0\times10^{-6}\times0.3166$ as shown in Tab.~\ref{tab:comoving}. That is to say, simulations with $\bar{\rho}_m^S*=6.0\times10^{-6}\times0.3166$ from $a^S=1$ to $a^S=1090$ can give the evolution of our universe with $\bar{\rho}_m^P*=6.0\times10^{-9}\times0.3166$ from $a^P=0.00092$ to $a^P=1$ when we analyze the results from simulations taking this blowing-up by 109 times into consideration and regardless of the existence of dark energy and radiation.
\begin{table*}[!htp]
\centering
\renewcommand{\arraystretch}{1.5}
\begin{tabular}{cccccccc}
\hline
  $\Omega_b h^2$ & $\Omega_c h^2$ & $\Omega_m$ & $H_0[\text{km}~\text{s}^{-1}\text{Mpc}^{-1}]$ & $n_s$ & $10^9A_s$ & $z_*$ & $z_{\text{re}}$ \\
\hline
$0.02236$ & $0.1202$ & $0.3166$ & $67.27$ & $0.9649$ & $2.101$ & $1089$ & $7.68$ \\
\hline
\end{tabular}
\caption{The cosmological parameters predicted by Planck 2018 TT,TE,EE+lowE \cite{Aghanim:2018eyx}.}
\label{tab:parameter}
\end{table*}
\begin{table*}[!htp]
\centering
\renewcommand{\arraystretch}{1.5}
\begin{tabular}{cccc}
\hline
   & $z+1=1090$ & $z+1=10$ & $z+1=1$ \\
\hline
$a^P$ & $0.00092$ & $0.1$ & $1$ \\
\hline
$a^F$ & $0.0092$ & $1$ & $10$  \\
\hline
$a^S$ & $1$ & $109$ & $1090$  \\
\hline
\end{tabular}
\caption{Three conversions between scale factor and redshift $z$. $a^P$ follows the usual convention in cosmology. $a^S$ is used during our simulations. $a^F$ is a fiducial one which relates the former two.}
\label{tab:a}
\end{table*}
\begin{table*}[!htp]
\centering
\renewcommand{\arraystretch}{1.5}
\begin{tabular}{cccc}
\hline
   & $a^P=1$  & $a^F=1$ & $a^S=1$ \\
\hline
$\bar{\rho}_m^I*$~~& $6.0\times10^{-9}\times0.3166$ ~~& $6.0\times10^{-6}\times0.3166$ ~~& $6.0\times10^{-6}\times0.3166$  \\
\hline
\end{tabular}
\caption{The comoving matter density $\bar{\rho}_m^I*=\bar{\rho}_m^I(a^I)^3$ for three different universes, where $I=P, F, S$ for our universe, fiducial universe and simulations respectively.}
\label{tab:comoving}
\end{table*}

All in all, we set the initial scale factor and matter background density for simulations as $a^S_\text{init}=1$ and $\bar{\rho}_{m,\text{init}}^S=6.0\times10^{-6}\times0.3166$ respectively.

\subsection{Initial conditions for perturbations}
\label{perturbations}

In the conformal Newtonian gauge, the line element that includes both the scalar and tensor perturbations to a spatially flat FLRW background spacetime is
\begin{equation}
\label{metric}
ds^2=(a^S)^2[-(1+2\Psi)d\eta^2+(1-2\Phi)\delta_{ij}dx^idx^j+h_{ij}dx^idx^j],
\end{equation}
where $\eta$ is the conformal time, $\delta_{ij}$ is the identity matrix, $\Psi$ is the Newtonian potential, $\Phi$ the spatial curvature perturbation and $h_{ij}$ is a divergenceless, traceless and symmetric tensor.
At the beginning of simulations, it's reasonable to take (\ref{metric}) as the universe's metric and rewrite it into the form of $(3+1)$ formalism
\begin{equation}
ds^2=-\alpha^2dt^2+\gamma_{ij}(dx^i+\beta^idt)(dx^j+\beta^jdt),
\end{equation}
where $\alpha$ is the lapse function which satisfies the harmonic slicing here: $\partial_t\alpha=-\frac{1}{3}\alpha^2K$, $\beta^i$ is the shift vector which is set as $\beta^i=0$ here and $\gamma_{ij}$ is the spatial metric which evolves depending on the extrinsic curvature $K_{ij}$ as
$(\partial_t-\mathcal{L}_{\vec{\beta}})\gamma_{ij}=-2\alpha K_{ij}$.
Therefore, the initial data for thorn \textsf{ADMBase} and \textsf{HydroBase} can be derived from the solutions at $\eta=0$ to Einstein's equation for (\ref{metric}).

Given the energy-momentum tensor of a perfect fluid without the anisotropic stress tensor $T_{\mu\nu}=(\rho+P)u_\mu u_\nu+Pg_{\mu\nu}$, we can give the evolutions of $a^S$ and  $\bar{\rho}_m^S$ according to the dust ($P\ll\rho\equiv\bar{\rho}^S_m(1+\delta)$) solutions to the zero-order Einstein equations for (\ref{metric})
\begin{eqnarray}
\label{sovleback}
a^S&=&a^S_{\mathrm{init}}\xi^2,\\\nonumber
\bar{\rho}^S_m&=&\bar{\rho}^S_{m,\text{init}}\xi^{-6},\\\nonumber
\xi&=&1+\sqrt{\frac{2\pi\bar{\rho}^S_m*}{3a^S_{\mathrm{init}}}}\eta,\\\nonumber
\xi&=&\left(\sqrt{6\pi\bar{\rho}^S_{m,\text{init}}}\int\alpha(t)dt+1\right)^{1/3}.
\end{eqnarray}
It's obviously that $a^S$, $\bar{\rho}^S_m$, $\xi$ and $\eta$ are functions of $t$ for FLRW background spacetime and they will become space-dependent in an inhomogeneous spacetime. For the latter case, we still take them as background quantities by taking the average of them across the simulation box.
Also, we can give the evolutions of perturbations according to the solutions to first-order Einstein equations for (\ref{metric})
\begin{eqnarray}
\label{sovlescalar}
\Phi&=&\Psi=f(x^i),\\\nonumber
\delta&=&C_1\xi^2\nabla^2f(x^i)-2f(x^i),\\\nonumber
v^i&=&C_2\xi\partial^if(x^i),\\\nonumber
h_{ij}&=&\int \frac{d^3k}{(2\pi)^3}h^{s}_k(\eta)\varepsilon^s_{ij}e^{i\vec{k}\cdot\vec{x}},
\end{eqnarray}
where $f(x^i)$ is an arbitrary function of space, $C_1=\frac{a^S_{\text{init}}}{4\pi\bar{\rho}_m^S*}$, $C_2=-\sqrt{\frac{a^S_{\text{init}}}{6\pi\bar{\rho}_m^S*}}$, where $\varepsilon^s_{ij}$ with $s=\times,+$ are transverse and traceless polarization tensors and each of $h^{s}_k(\eta)$ evolves independently and satisfies $h_{k}^s(\eta+\eta_0)=3h_{k}^s(0)\frac{\sin[k(\eta+\eta_0)]-[k(\eta+\eta_0)]\cos[k(\eta+\eta_0)]}{[k(\eta+\eta_0)]^3}$.
According to (\ref{sovleback}) and (\ref{sovlescalar}), at $\eta=0$ (or $\xi=1$), the initial data will dependent on $a^S_{\mathrm{init}}$, $\bar{\rho}^S_{m,\text{init}}$, $f(x^i)$, $h_{k}^s(0)$ and $\eta_0$.

\begin{figure}[]
\begin{center}
\includegraphics[scale=0.35]{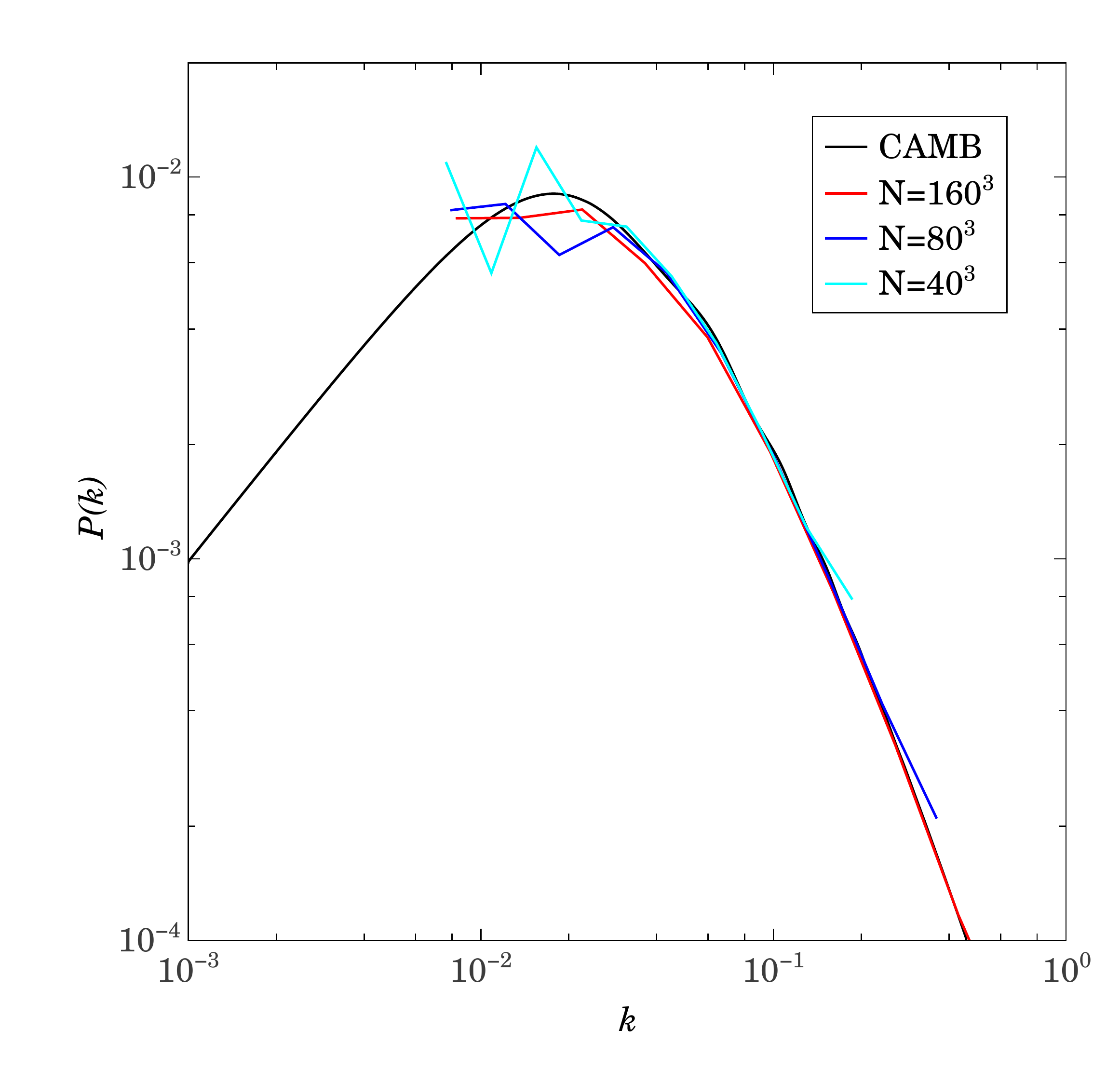}
\includegraphics[scale=0.35]{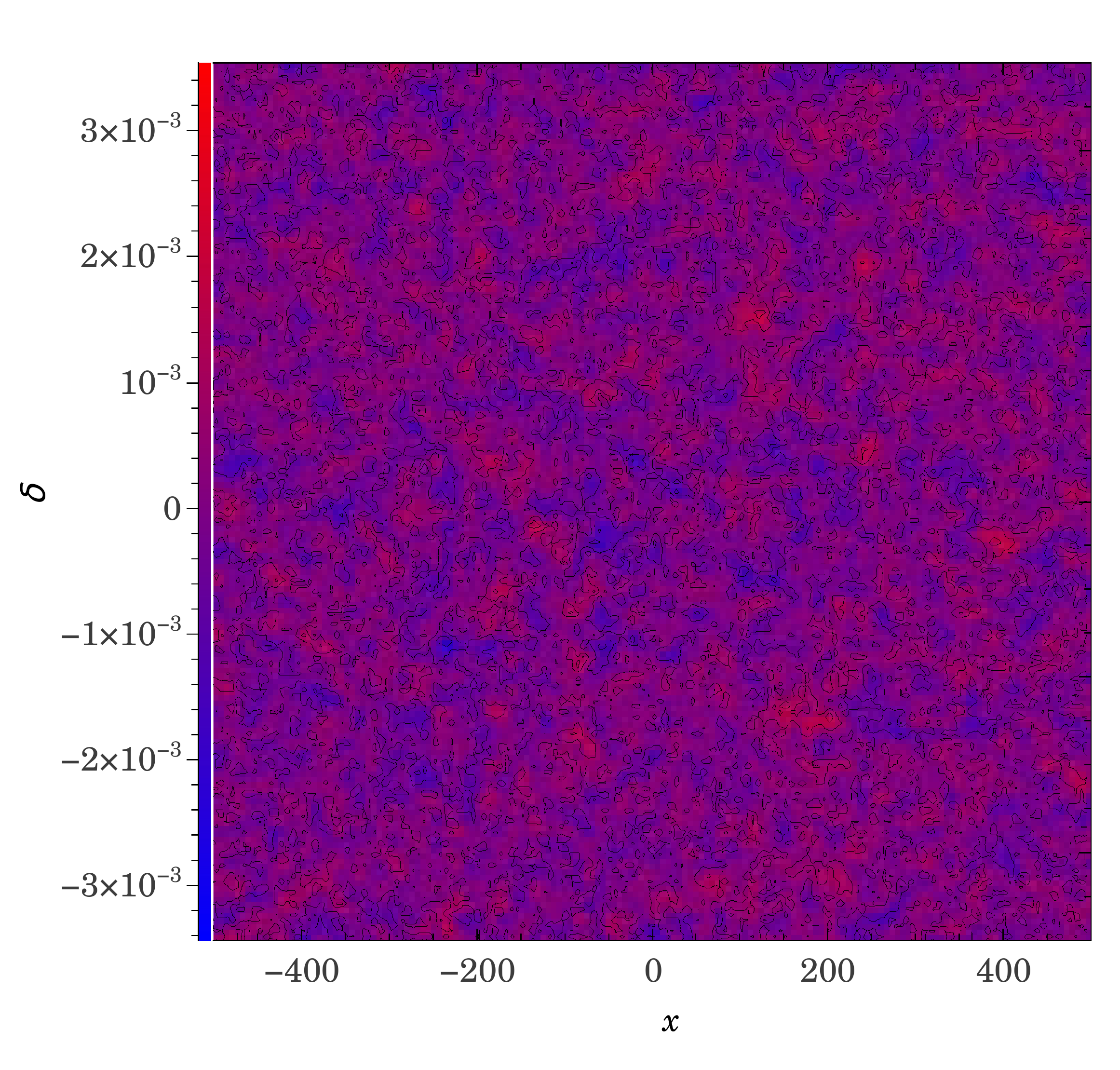}
\includegraphics[scale=0.35]{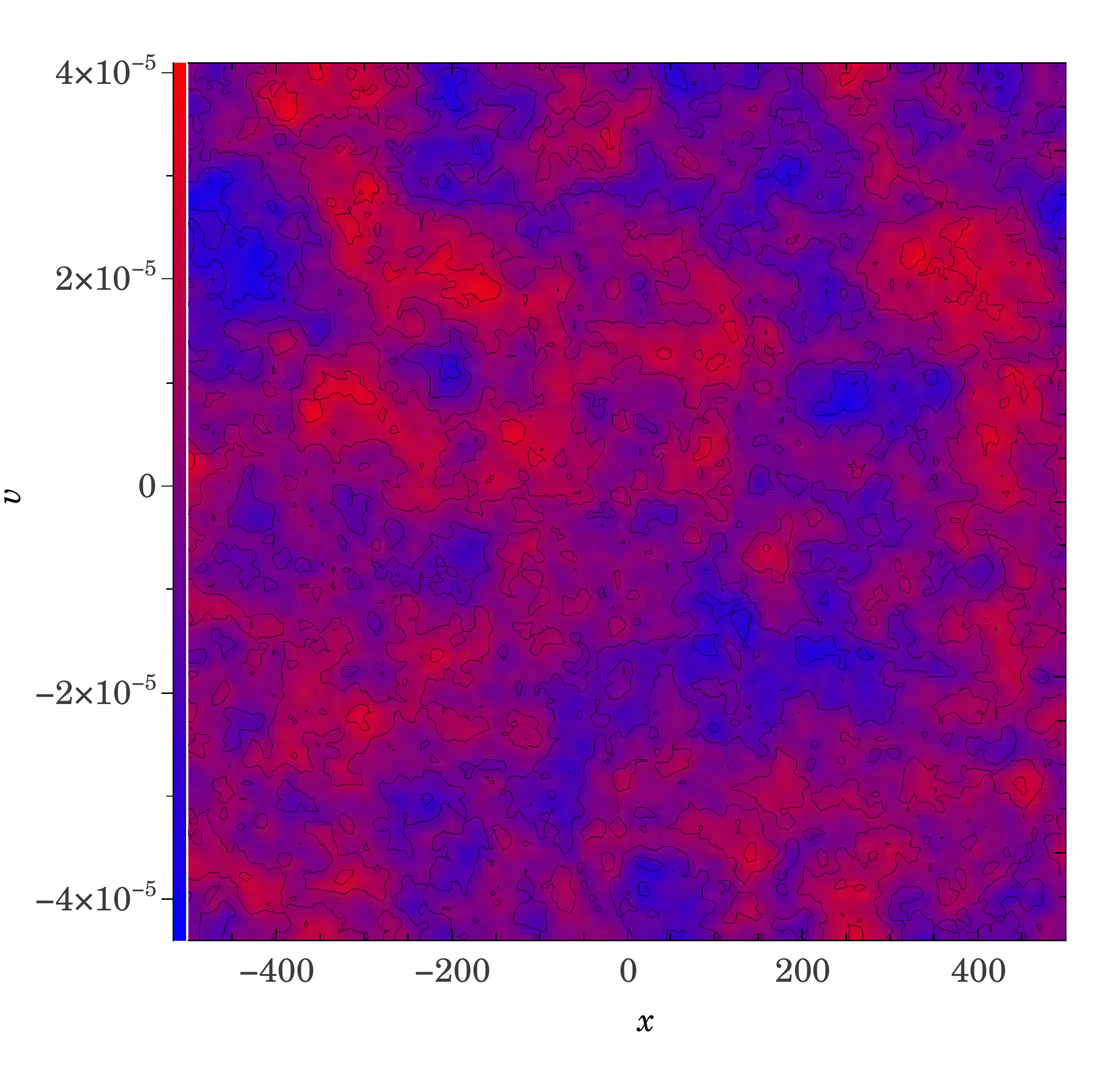}
\includegraphics[scale=0.35]{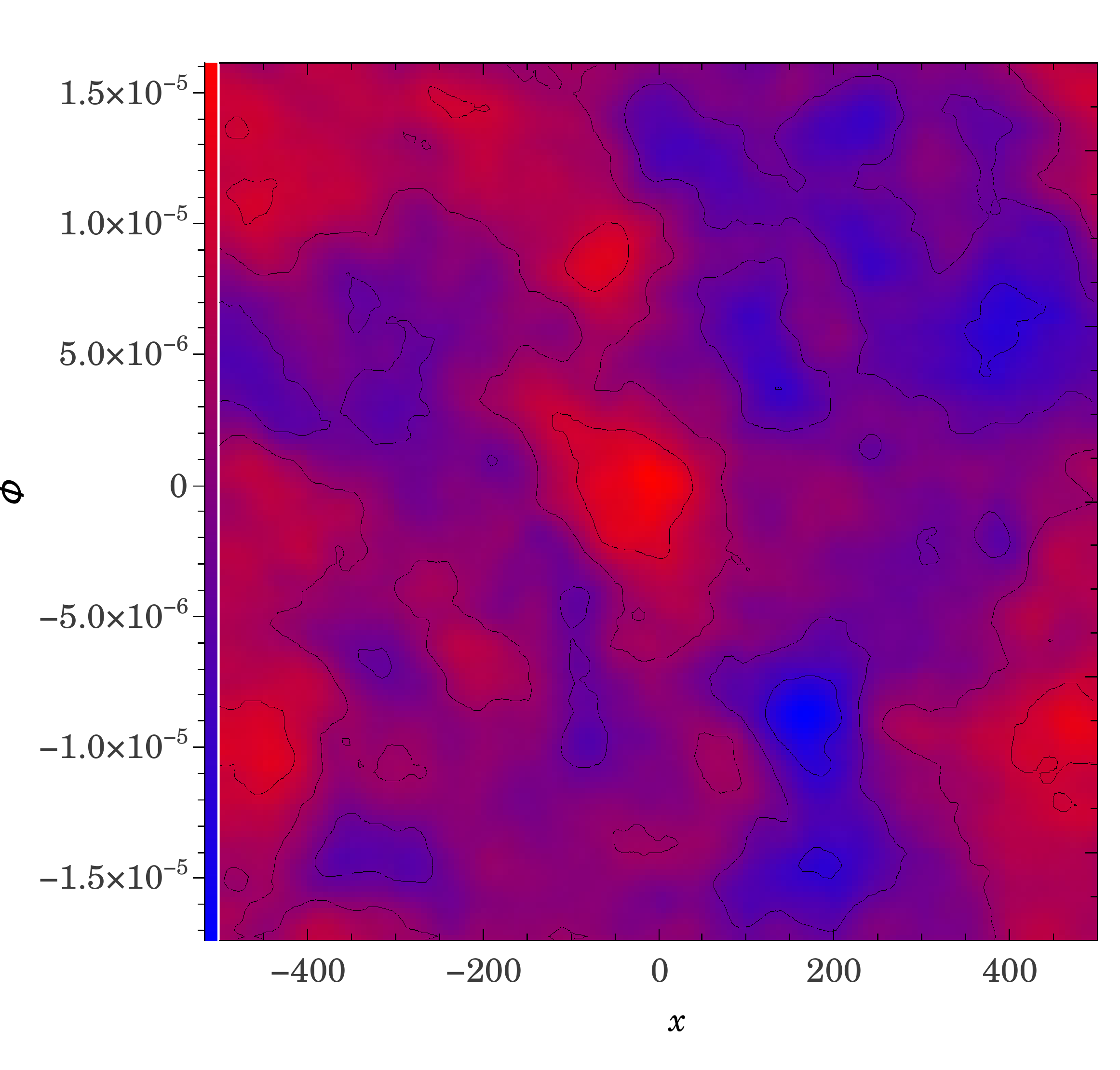}
\end{center}
\caption{Matter power spectrum and the initial conditions derived from it. There are four curves in the first plot: the black one is the matter power spectrum at $z+1=1090$ produced by CAMB \cite{Lewis:2002ah} with parameters listed in Tab.~\ref{tab:parameter}; the red, blue and cyan curves are the matter power spectra drawn from density perturbations $\delta(x^i)$ by the function \textsf{power\_spectrum\_1d} in \textsf{c2raytools}\cite{c2ray} at $160^3$, $80^3$ and $40^3$ resolution respectively, while $\delta(x^i)$ is generated by the function \textsf{make\_gaussian\_random\_field} in \textsf{c2raytools} from the black curve. The second plot is the distribution of $\delta(x^i)$ at $a^S=1$ and $160^3$ resolution. The last two plots are the distribution of $v^x(x^i)$ and $\Phi(x^i)$ at $a^S=1$ respectively, which are derived from $\delta(x^i)$ according to the Fourier version of (\ref{sovlescalar}). }
\label{fig:intial}
\end{figure}

The last plot in Fig.~\ref{fig:intial} shows the distribution of spatial curvature perturbations $\Phi(x^i)$ (or $f(x^i)$) at $a^S=1$. In fact, we use the function \textsf{make\_gaussian\_random\_field} in \textsf{c2raytools} \cite{c2ray} to generate the density perturbations $\delta(x^i)$ (the second plot in Fig.~\ref{fig:intial}) from the matter power spectrum at $z+1=1090$ (the first plot in Fig.~\ref{fig:intial}) produced by the Code for Anisotropies in the Microwave Background (CAMB) \cite{Lewis:2002ah} with parameters listed in Tab.~\ref{tab:parameter} first. And then we derive $f(x^i)$ from $\delta(x^i)$ according to the Fourier version of (\ref{sovlescalar}), hence $\Phi(x^i)$, $\Psi(x^i)$ and $v^i(x^i)$ (the third plot in Fig.~\ref{fig:intial}). As for tensor perturbations, we here only consider one single mode with $k=\frac{2\pi}{L}$ and the space distribution as $\cos(\frac{2\pi}{L} z)$, where $L=1000$ is the length of one side of our simulation box with $x^i$ in $[-500, 500]$. And we set its initial amplitude $h_{\frac{2\pi}{L}}^s(0)=10^{-3}$, but it has crossed inside the horizon and decayed by $70\%$ when $\eta_0\simeq2\sqrt{\frac{3}{8\pi\bar{\rho}^S_{m,\text{init}}}}$.

\section{Results}
\label{results}
\begin{figure}[]
\begin{center}
\includegraphics[scale=0.4]{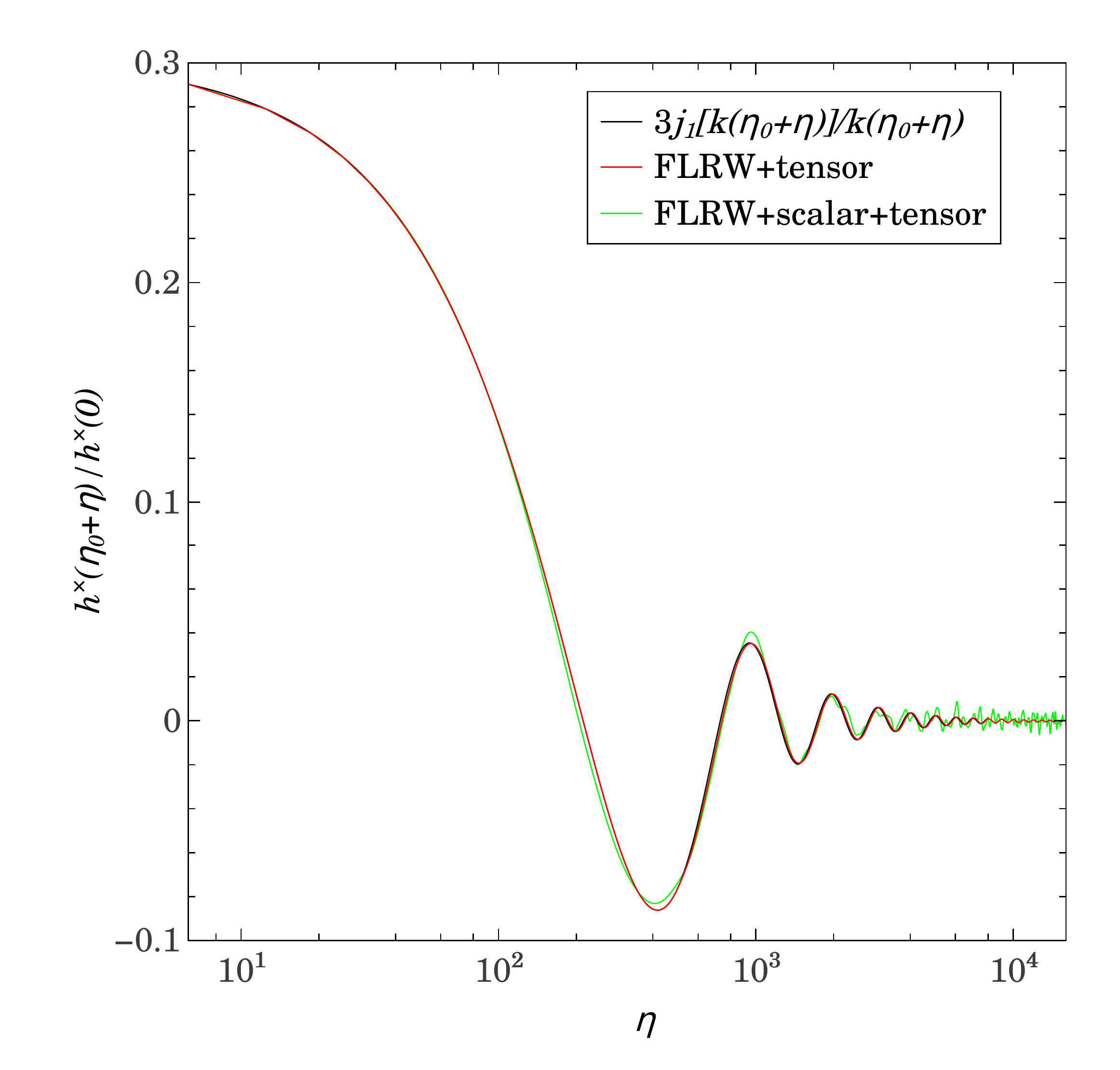}
\end{center}
\caption{The evolution of $\frac{h^{\times}(\eta_0+\eta)}{h^{\times}(0)}$ at the origin of our simulation box. The black curve is drawn by $\frac{3j_1[k(\eta_0+\eta)]}{k(\eta_0+\eta)}$, where $j_1(z)=\frac{\sin z-z\cos z}{z^2}$ is the spherical Bessel functions of order one. The red curve is the evolution of $\frac{\gamma_{12}(\eta)}{(a^S)^2}$ in simulations with only tensor perturbations. The green curve is the evolution of $\frac{\gamma_{12}(\eta)}{(a^S)^2}$ in simulations with scalar and tensor perturbations. We can see that the black curve and the red one are almost coincide and there are slight deviations between the red curve and the green one. That is to say, we can relate $\frac{\gamma_{12}(\eta)}{(a^S)^2}$ to the evolution of tensor perturbation $h^{\times}(\eta_0+\eta)$ in our simulations. }
\label{fig:hx}
\end{figure}

\begin{figure}[]
\begin{center}
\includegraphics[scale=0.35]{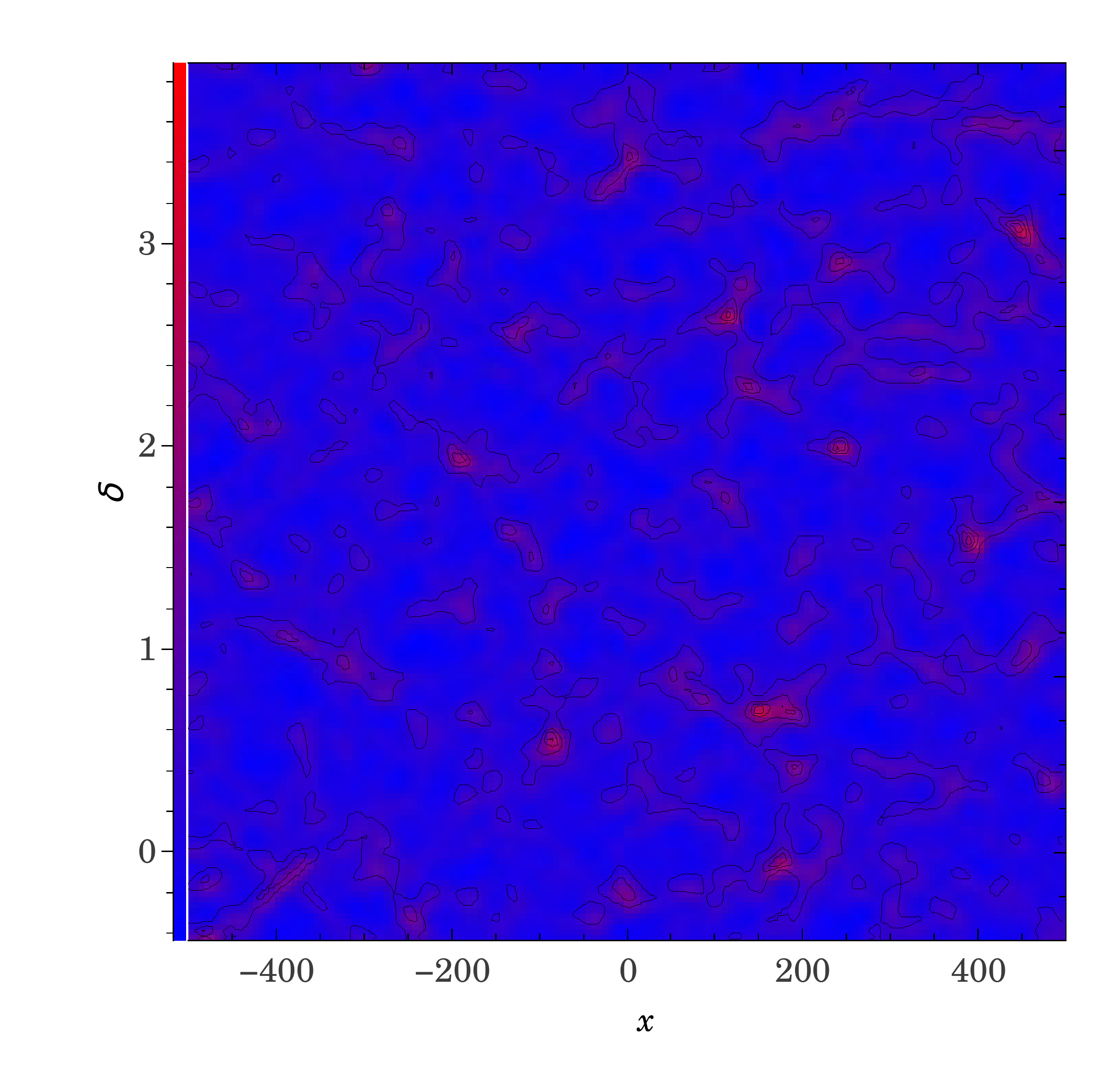}
\includegraphics[scale=0.35]{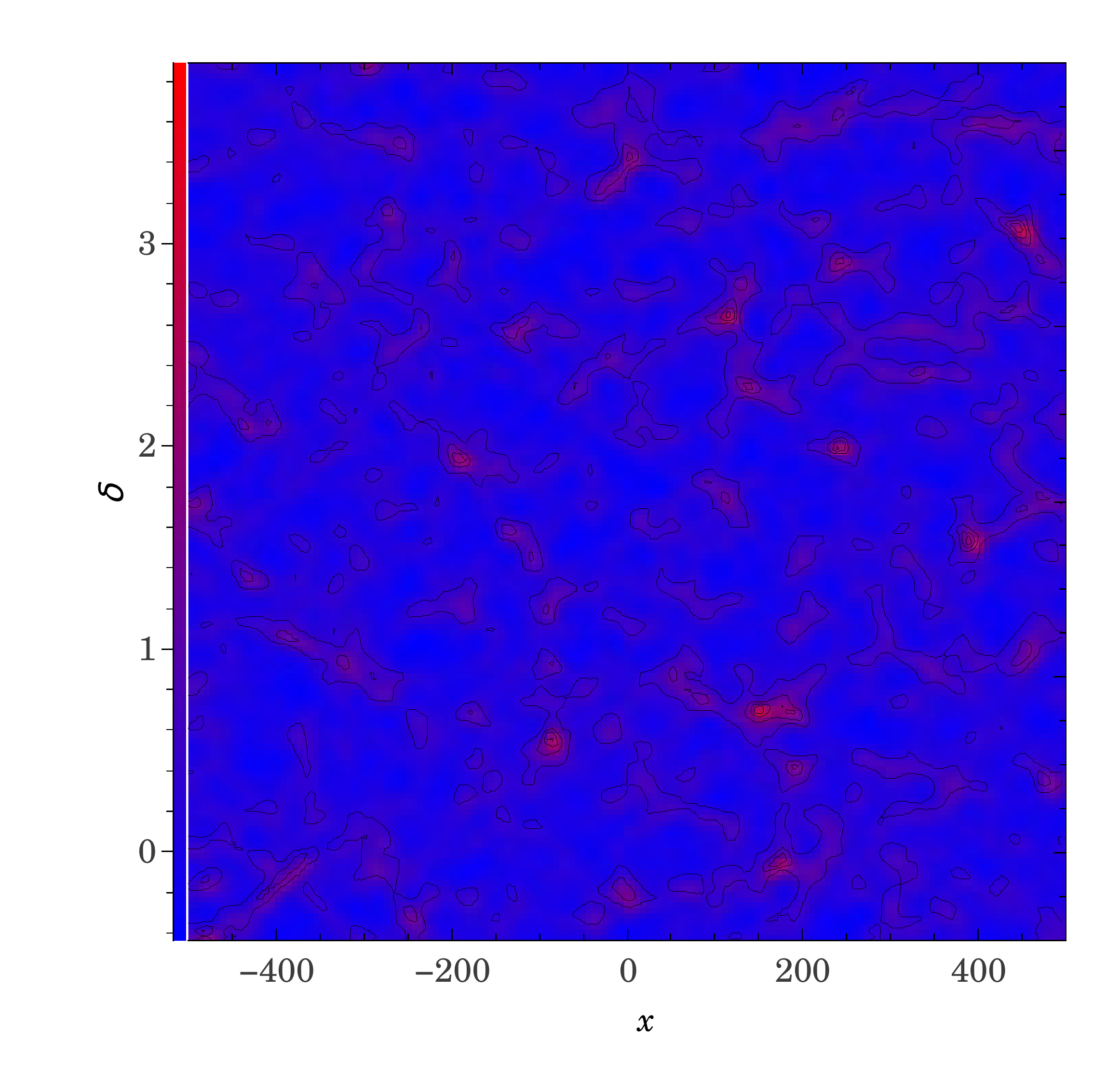}
\end{center}
\caption{The distribution of $\delta(x^i)$ without tensor perturbations (left) and with tensor perturbations (right) at $a^S=1090$ and $160^3$ resolution. It's hard to distinguish the effects of tensor perturbations from them directly. Therefore, we will turn to the matter power spectrum here. Moreover we can see there is a nonlinear web structure.}
\label{fig:finaldelta}
\end{figure}

\begin{figure}[]
\begin{center}
\includegraphics[scale=0.35]{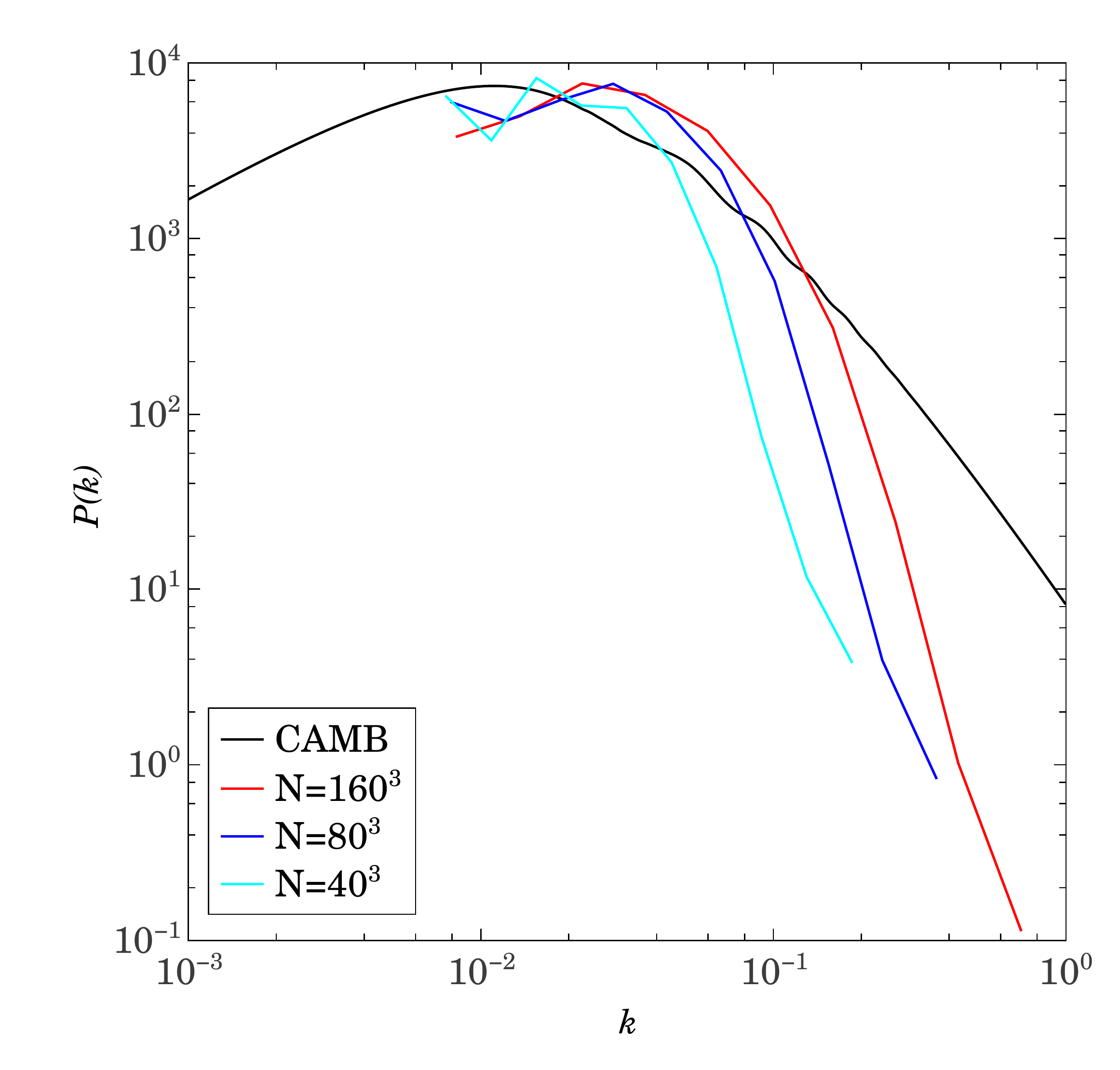}
\includegraphics[scale=0.35]{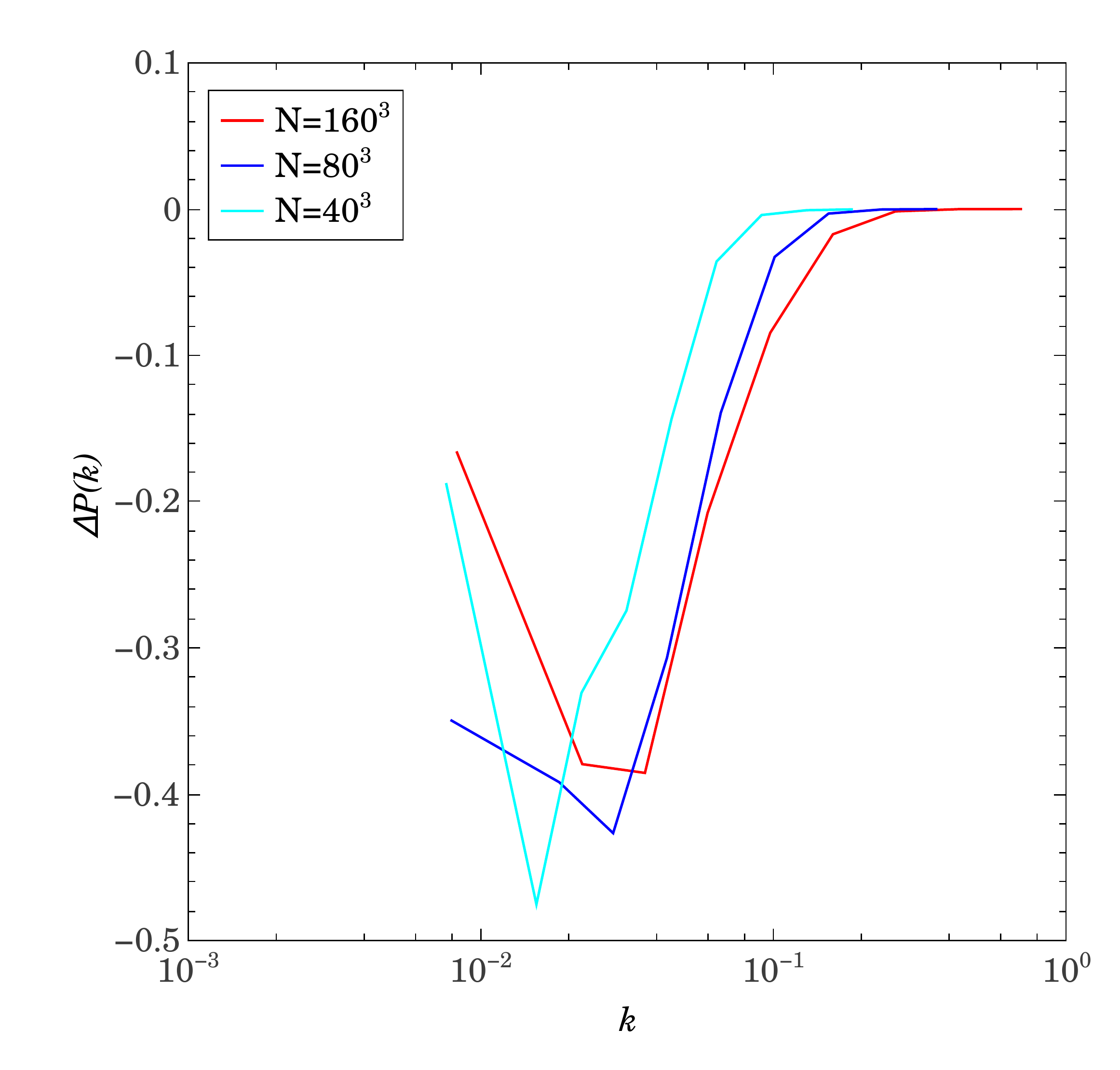}
\end{center}
\caption{Matter power spectra and the effects of tensor perturbations on spectra. The black curve in the left plot is the linear matter power spectrum at $z+1=1$ produced by CAMB with parameters listed in Tab.~\ref{tab:parameter}; the red, blue and cyan curves in the left plot are the matter power spectra drawn from density perturbations $\delta(x^i)$ at $a^S=1090$ by the function \textsf{power\_spectrum\_1d} in \textsf{c2raytools} at $160^3$, $80^3$ and $40^3$ resolution respectively, where $\delta(x^i)$ is given by simulations with only scalar perturbations. It's worth pointing out that although, at the beginning of simulations, our initial data is derived from matter power spectrum at $z+1=1090$, we ignore the dark energy in our simulations at the late time. So the black curve has a different trend with color ones in the left plot. The red, blue and cyan curves in the right plot explicitly show the suppression of matter power spectra for modes with wave number similar to the tensor perturbations' at $160^3$, $80^3$ and $40^3$ resolution respectively. And this suppression converge to about $0.01\%$. }
\label{fig:pk}
\end{figure}

Our simulations are performed at $160^3$, $80^3$ and $40^3$ resolution and end at $a^S=1090$. Due to the coincidence of the black curve drawn by $\frac{3j_1[k(\eta_0+\eta)]}{k(\eta_0+\eta)}$ (where $j_1(z)=\frac{\sin z-z\cos z}{z^2}$ is the spherical Bessel functions of order one) and the red one which is the evolution of $\frac{\gamma_{12}(\eta)}{(a^S)^2}$ given by simulations with only tensor perturbations, in Fig.~\ref{fig:hx}, we relate $\frac{\gamma_{12}(\eta)}{(a^S)^2}$ to the evolution of tensor perturbation $h^{\times}(\eta_0+\eta)$ in our simulations. Although, as shown in Fig.~\ref{fig:hx}, there are slight deviations between the red curve and the green one which is the evolution of $\frac{\gamma_{12}(\eta)}{(a^S)^2}$ given by simulations with scalar and tensor perturbations, we keep this relation standing. For probing the effects of primordial tensor perturbations on the inhomogeneity of the universe, it's naive to compare the distribution of $\delta(x^i)$ at $a^S=1090$ given by simulations with scalar and tensor perturbations and their counterparts with only scalar perturbations directly, as shown in Fig.~\ref{fig:finaldelta}. Here we will turn to the the matter power spectrum, which is an important statistical quantity and can be detected by many experiments \cite{Reid:2009xm,Parkinson:2012vd,lsst,Abell:2009aa}. In the left plot of Fig.~\ref{fig:pk}, the red, blue and cyan curves are the matter power spectra drawn from density perturbations $\delta(x^i)$ at $a^S=1090$ by the function \textsf{power\_spectrum\_1d} in \textsf{c2raytools} at $160^3$, $80^3$ and $40^3$ resolution respectively, where $\delta(x^i)$ is given by simulations with only scalar perturbations. When taking the tensor perturbations into consideration, we can get similar matter power spectra. Comparing them with the formers, we can find an obvious suppression of matter power spectra for modes with wave number similar to the tensor perturbations', as shown in the right plot of Fig.~\ref{fig:pk}. And comparing the suppression at $160^3$, $80^3$ and $40^3$ resolution, we can find this suppression converge to about $0.01\%$ if the initial amplitude of the tensor perturbations is $3\times 10^{-4}$.

\begin{figure}[]
\begin{center}
\includegraphics[scale=0.35]{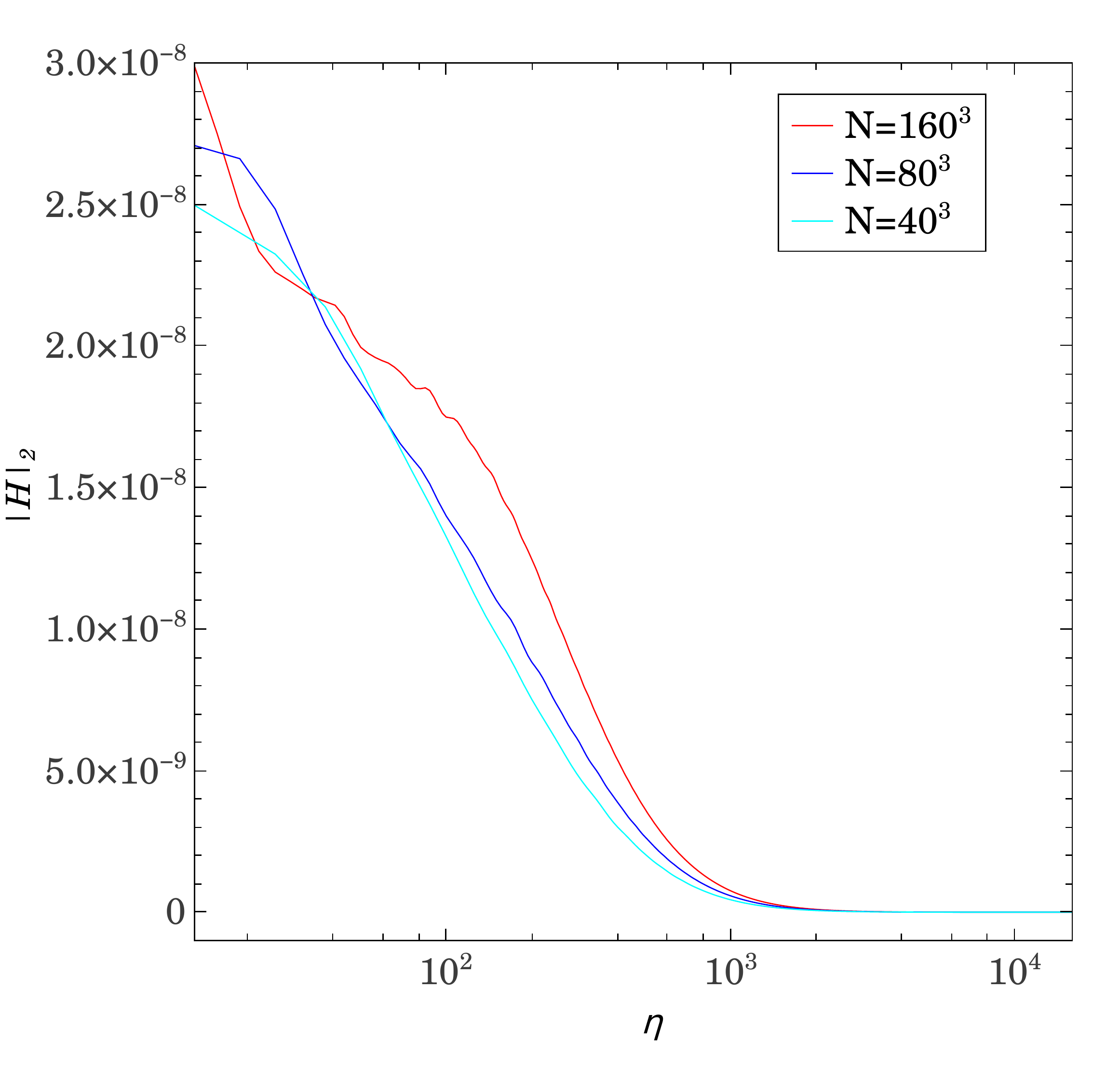}
\includegraphics[scale=0.35]{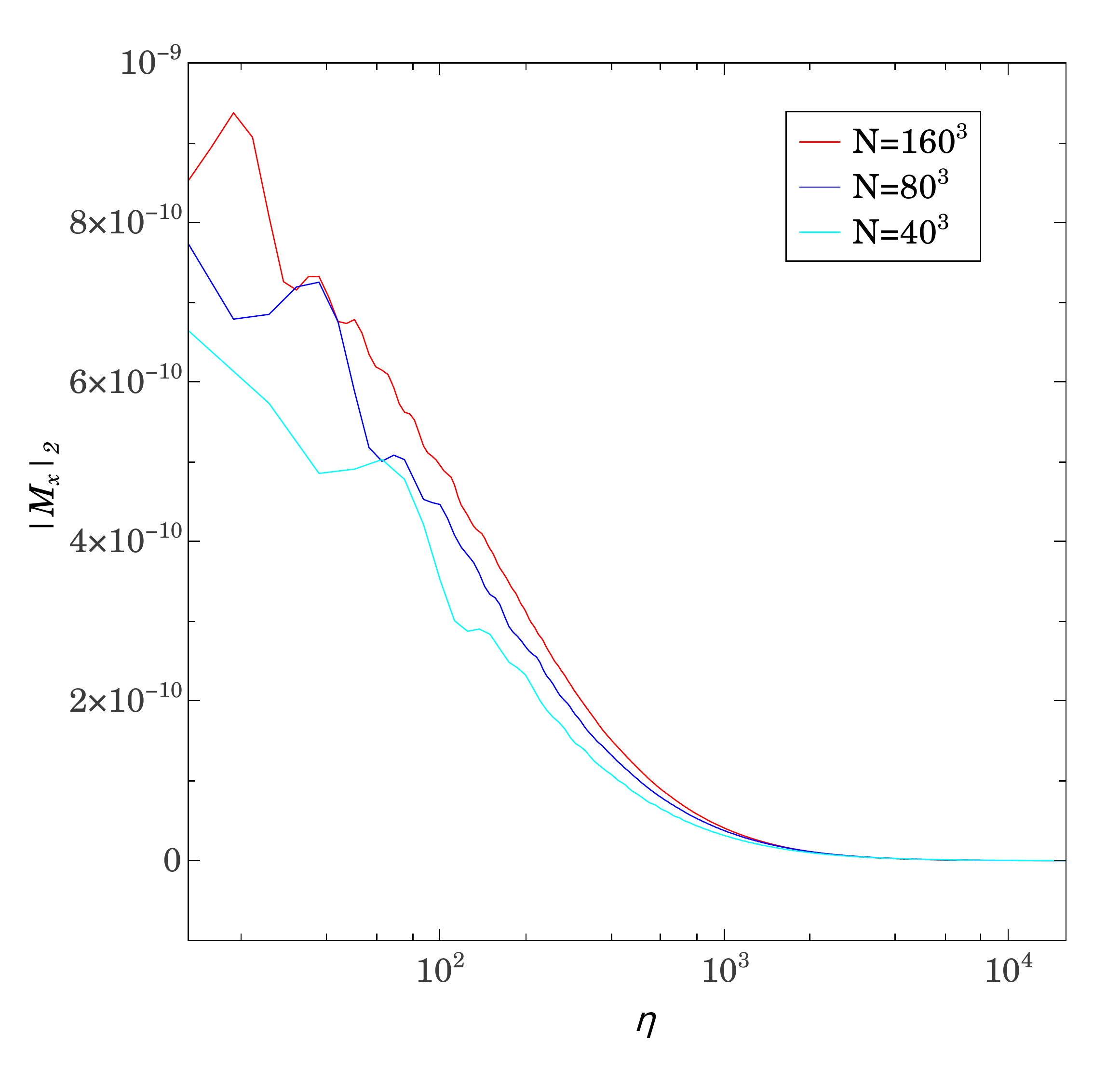}
\end{center}
\caption{$L_2$ norms of the Hamiltonian constraint violation and the x-component of momentum constraint violation at $160^3$ (red), $80^3$ (blue) and $40^3$ (cyan) resolution.}
\label{fig:constraint}
\end{figure}

Even though the initial conditions derived from the matter power spectrum at $z+1=1090$ satisfy the perturbed Einstein equations, it's still necessary to check that to what extend do these initial data satisfy the Hamiltonian constraint and the momentum constraint. Given the 3-Riemann scalar $^{(3)}R$, the covariant derivative associated with the 3-metric $D_j$, and the matter energy and momentum density as measured by the Eulerian observer $E$ and $p_i$, we can specify the form of the Hamiltonian constraint violation and the momentum constraint violation as
\begin{equation}
\mathcal{H}=\frac{1}{2}(^{(3)}R+K^2-K_{ij}K^{ij})-8\pi E
\end{equation}
and
\begin{equation}
\mathcal{M}_i=D_jK^j_i-D_iK-8\pi p_i.
\end{equation}
Fig.~\ref{fig:constraint} shows the evolution of $L_2$ norms of the Hamiltonian constraint violation and the x-component of momentum constraint violation at $160^3$, $80^3$ and $40^3$ resolution. We can see that the higher resolution, the larger constraint violation. The reason for this abnormal behaviour is that the initial $\delta(x^i)$ generated by the function \textsf{make\_gaussian\_random\_field} in \textsf{c2raytools} from the matter power spectrum at $z+1=1090$ produced by CAMB is resolution-dependent: the higher resolution leads to $\delta(x^i)$ with larger wave number; the scalar perturbations on smaller scales have larger amplitude.  As pointed out in \cite{Macpherson:2018btl}, one can present the convergence of constraint violation explicitly by transferring raw constraint violation to relative one.

\section{Summary and discussion}
\label{summary}
We simulate a dust universe from $a^S=1$ (or $z=1089$) to $a^S=1090$ (or $z=0$) by numerically integrating the Einstein's equation whose solution at $a^S=1$ is a spatially flat FLRW metric with scalar perturbations which are derived from the matter power spectrum produced with CAMB. Then we add an additional decaying, divergenceless and traceless primordial tensor perturbation with its initial amplitude being $3\times 10^{-4}$ to the metric as shown in Fig.~\ref{fig:hx}. Simulations at $160^3$, $80^3$ and $40^3$ resolution converge and show that this primordial tensor perturbation suppresses the matter power spectrum by about $0.01\%$ at $z=0$ for modes with wave number $k\sim0.05$ as shown in Fig.~\ref{fig:pk}.

In the linear perturbation theory, scalar and tensor perturbations are supposed to be totally decoupled. 
However, there are some non-linear coupling terms between scalar and tensor perturbations for the full Einstein equations which are used in our simulations. 
Even though we turn to the first-order perturbed Einstein equations for the initial data, they satisfy the full Einstein constraints of early universe just with tiny deviations.
That is to say, \textsf{Einstein Toolkit} takes the all possible terms of the full Einstein equations into our consideration. 
Therefore, this suppression results from the fully relativistic treatment for Einstein equations. Although there are nonlinear structures formed at the end of simulations ($a^S=1090$) as shown in Fig.~\ref{fig:finaldelta}, their scales are smaller than tensor perturbations'. So this suppression sown before the tensor perturbations died out and amplified with time is still in linear regime.

There are two caveats. First the production of monochromatic single mode gravitational wave seems unrealistic in cosmology and most inflation models predict a scale-invariant spectrum of gravitational waves. Here we only consider a monochromatic gravitational wave because primordial gravitational waves enter the horizon one by one. Given the comoving length of one side of our simulation box $L=1000$ and the initial matter density, the modes with wave length $>1000$ are initially outside the simulation box and will never enter it during simulations. As for modes with wave length $<1000$, they entered the horizon earlier and almost died out. Therefore, if we want to study scale dependence of the results, we must perform simulations under other larger $L$, which results in high computational cost. Here we just make our results as a first step to more comprehensive studies. Also it's necessary to include the dark energy if one want to compare the results with observations. Because dark energy is supposed to affect the very late-time growth factor by about $10\%$. So far, however, people can't simulate dark energy in \textsf{Einstein Toolkit}. Here we just keep it in mind.

This suppression may be a possible probe of a GWs background in the future only if the matter power spectrum is measured in high enough precision. Undoubtedly, by the time LSST is in full operation, the required precision for detection of such suppression is still far beyond reach. However, this suppression is an unique signature put by primordial GWs.

\vspace{5mm}
\noindent {\bf Acknowledgments}
We would like to thank You-Jun Lu for his helpful discussions and advices on this paper. This work is partly supported by the National Natural Science Foundation of China under grant No. 11690024, the Strategic Priority Program of the Chinese Academy of Sciences (Grant No. XDB 23040100).


\newpage




\begin{thebibliography}{99}
\frenchspacing

\bibitem{Starobinsky:1980te}
  A.~A.~Starobinsky,
  Phys.\ Lett.\  {\bf 91B}, 99 (1980).
  doi:10.1016/0370-2693(80)90670-X

\bibitem{Guth:1980zm}
  A.~H.~Guth,
  Phys.\ Rev.\ D {\bf 23}, 347 (1981).
  doi:10.1103/PhysRevD.23.347

\bibitem{Seljak:1996gy}
  U.~Seljak and M.~Zaldarriaga,
  Phys.\ Rev.\ Lett.\  {\bf 78}, 2054 (1997)
  doi:10.1103/PhysRevLett.78.2054
  [astro-ph/9609169].

\bibitem{Kamionkowski:1996zd}
  M.~Kamionkowski, A.~Kosowsky and A.~Stebbins,
  Phys.\ Rev.\ Lett.\  {\bf 78}, 2058 (1997)
  doi:10.1103/PhysRevLett.78.2058
  [astro-ph/9609132].

\bibitem{Kamionkowski:2015yta}
  M.~Kamionkowski and E.~D.~Kovetz,
  Ann.\ Rev.\ Astron.\ Astrophys.\  {\bf 54}, 227 (2016)
  doi:10.1146/annurev-astro-081915-023433
  [arXiv:1510.06042 [astro-ph.CO]].

\bibitem{Book:2011dz}
  L.~Book, M.~Kamionkowski and F.~Schmidt,
  Phys.\ Rev.\ Lett.\  {\bf 108}, 211301 (2012)
  doi:10.1103/PhysRevLett.108.211301
  [arXiv:1112.0567 [astro-ph.CO]].

\bibitem{Masui:2010cz}
  K.~W.~Masui and U.~L.~Pen,
  Phys.\ Rev.\ Lett.\  {\bf 105}, 161302 (2010)
  doi:10.1103/PhysRevLett.105.161302
  [arXiv:1006.4181 [astro-ph.CO]].

\bibitem{Dodelson:2003bv}
  S.~Dodelson, E.~Rozo and A.~Stebbins,
  Phys.\ Rev.\ Lett.\  {\bf 91}, 021301 (2003)
  doi:10.1103/PhysRevLett.91.021301
  [astro-ph/0301177].

\bibitem{Dodelson:2010qu}
  S.~Dodelson,
  Phys.\ Rev.\ D {\bf 82}, 023522 (2010)
  doi:10.1103/PhysRevD.82.023522
  [arXiv:1001.5012 [astro-ph.CO]].

\bibitem{Jeong:2012nu}
  D.~Jeong and F.~Schmidt,
  Phys.\ Rev.\ D {\bf 86}, 083512 (2012)
  doi:10.1103/PhysRevD.86.083512
  [arXiv:1205.1512 [astro-ph.CO]].

\bibitem{Schmidt:2012nw}
  F.~Schmidt and D.~Jeong,
  Phys.\ Rev.\ D {\bf 86}, 083513 (2012)
  doi:10.1103/PhysRevD.86.083513
  [arXiv:1205.1514 [astro-ph.CO]].

\bibitem{Hu:1997mj}
  W.~Hu, D.~J.~Eisenstein and M.~Tegmark,
  Phys.\ Rev.\ Lett.\  {\bf 80}, 5255 (1998)
  doi:10.1103/PhysRevLett.80.5255
  [astro-ph/9712057].

\bibitem{Lesgourgues:2006nd}
  J.~Lesgourgues and S.~Pastor,
  Phys.\ Rept.\  {\bf 429}, 307 (2006)
  doi:10.1016/j.physrep.2006.04.001
  [astro-ph/0603494].

\bibitem{Riemer-Sorensen:2013jsa}
  S.~Riemer-Sørensen, D.~Parkinson and T.~M.~Davis,
  Phys.\ Rev.\ D {\bf 89}, 103505 (2014)
  doi:10.1103/PhysRevD.89.103505
  [arXiv:1306.4153 [astro-ph.CO]].

\bibitem{Palanque-Delabrouille:2015pga}
  N.~Palanque-Delabrouille {\it et al.},
  JCAP {\bf 1511}, no. 11, 011 (2015)
  doi:10.1088/1475-7516/2015/11/011
  [arXiv:1506.05976 [astro-ph.CO]].

\bibitem{Cuesta:2015iho}
  A.~J.~Cuesta, V.~Niro and L.~Verde,
  Phys.\ Dark Univ.\  {\bf 13}, 77 (2016)
  doi:10.1016/j.dark.2016.04.005
  [arXiv:1511.05983 [astro-ph.CO]].

\bibitem{Reid:2009xm}
  B.~A.~Reid {\it et al.},
  Mon.\ Not.\ Roy.\ Astron.\ Soc.\  {\bf 404}, 60 (2010)
  doi:10.1111/j.1365-2966.2010.16276.x
  [arXiv:0907.1659 [astro-ph.CO]].

\bibitem{Parkinson:2012vd}
  D.~Parkinson {\it et al.},
  Phys.\ Rev.\ D {\bf 86}, 103518 (2012)
  doi:10.1103/PhysRevD.86.103518
  [arXiv:1210.2130 [astro-ph.CO]].

\bibitem{lsst}
  www.lsst.org

\bibitem{Abell:2009aa}
  P.~A.~Abell {\it et al.} [LSST Science and LSST Project Collaborations],
  arXiv:0912.0201 [astro-ph.IM].

\bibitem{Loffler:2011ay}
  F.~Loffler {\it et al.},
  Class.\ Quant.\ Grav.\  {\bf 29}, 115001 (2012)
  doi:10.1088/0264-9381/29/11/115001
  [arXiv:1111.3344 [gr-qc]].

\bibitem{Brown:2008sb}
  J.~D.~Brown, P.~Diener, O.~Sarbach, E.~Schnetter and M.~Tiglio,
  Phys.\ Rev.\ D {\bf 79}, 044023 (2009)
  doi:10.1103/PhysRevD.79.044023
  [arXiv:0809.3533 [gr-qc]].

\bibitem{Reisswig:2010cd}
  C.~Reisswig, C.~D.~Ott, U.~Sperhake and E.~Schnetter,
  Phys.\ Rev.\ D {\bf 83}, 064008 (2011)
  doi:10.1103/PhysRevD.83.064008
  [arXiv:1012.0595 [gr-qc]].

\bibitem{code}
  McLachlan, a public BSSN code URL. http://www.cct.lsu.edu/~eschnett/    McLachlan/

\bibitem{Baumgarte:1998te}
  T.~W.~Baumgarte and S.~L.~Shapiro,
  Phys.\ Rev.\ D {\bf 59}, 024007 (1999)
  doi:10.1103/PhysRevD.59.024007
  [gr-qc/9810065].

\bibitem{Shibata:1995we}
  M.~Shibata and T.~Nakamura,
  Phys.\ Rev.\ D {\bf 52}, 5428 (1995).
  doi:10.1103/PhysRevD.52.5428

\bibitem{Alcubierre:2000xu}
  M.~Alcubierre {\it et al.},
  Phys.\ Rev.\ D {\bf 62}, 044034 (2000)
  doi:10.1103/PhysRevD.62.044034
  [gr-qc/0003071].

\bibitem{Moesta:2013dna}
  P.~Mösta {\it et al.},
  Class.\ Quant.\ Grav.\  {\bf 31}, 015005 (2014)
  doi:10.1088/0264-9381/31/1/015005
  [arXiv:1304.5544 [gr-qc]].

\bibitem{Baiotti:2004wn}
  L.~Baiotti, I.~Hawke, P.~J.~Montero, F.~Loffler, L.~Rezzolla, N.~Stergioulas, J.~A.~Font and E.~Seidel,
  Phys.\ Rev.\ D {\bf 71}, 024035 (2005)
  doi:10.1103/PhysRevD.71.024035
  [gr-qc/0403029].

\bibitem{Hawke:2005zw}
  I.~Hawke, F.~Loffler and A.~Nerozzi,
  Phys.\ Rev.\ D {\bf 71}, 104006 (2005)
  doi:10.1103/PhysRevD.71.104006
  [gr-qc/0501054].

\bibitem{Macpherson:2016ict}
  H.~J.~Macpherson, P.~D.~Lasky and D.~J.~Price,
  Phys.\ Rev.\ D {\bf 95}, no. 6, 064028 (2017)
  doi:10.1103/PhysRevD.95.064028
  [arXiv:1611.05447 [astro-ph.CO]].

\bibitem{Wang:2018qfr}
  K.~Wang,
  Eur.\ Phys.\ J.\ C {\bf 78}, no. 8, 629 (2018)
  doi:10.1140/epjc/s10052-018-6103-7
  [arXiv:1801.08362 [astro-ph.CO]].

\bibitem{Macpherson:2018btl}
  H.~J.~Macpherson, D.~J.~Price and P.~D.~Lasky,
  Phys.\ Rev.\ D {\bf 99}, no. 6, 063522 (2019)
  doi:10.1103/PhysRevD.99.063522
  [arXiv:1807.01711 [astro-ph.CO]].

\bibitem{Aghanim:2018eyx}
  N.~Aghanim {\it et al.} [Planck Collaboration],
  arXiv:1807.06209 [astro-ph.CO].

\bibitem{c2ray}
  https://github.com/hjens/c2raytools

\bibitem{Lewis:2002ah}
  A.~Lewis and S.~Bridle,
  Phys.\ Rev.\ D {\bf 66}, 103511 (2002)
  doi:10.1103/PhysRevD.66.103511
  [astro-ph/0205436].





\end{thebibliography}
\end{document}